\documentclass[prl,aps,twocolumn,preprintnumbers,amsmath,amssymb,nofootinbib,superscriptaddress,notitlepage]{revtex4-1}

\usepackage{epsfig}
\usepackage[utf8]{inputenc}
\usepackage{color}
\usepackage{epstopdf}
\usepackage{multirow}
\usepackage{verbatim}
\usepackage{ftnxtra}

\newcommand{\be}{\begin{eqnarray}}
\newcommand{\ee}{\end{eqnarray}}

\newcommand{\uestc}{\affiliation{School of Physics, University of Electronic Science and
Technology of China, Chengdu 610054, China}} 
\newcommand{\ucas}{\affiliation{School of Physical Sciences, University of Chinese Academy of Sciences (UCAS), Beijing 100049, China}}
\newcommand{\asrc}{\affiliation{Advanced Science Research Center, Japan Atomic Energy
Agency, Tokai, Ibaraki, 319-1195, Japan}}

\newcommand{\riken}{\affiliation{Nishina Center
for Accelerator-Based Science, RIKEN, Wako 351-0198, Japan}}

\newcommand{\pku}{\affiliation{School of Physics and Center of High Energy Physics,
Peking University, Beijing 100871,China}}

\begin{document}

\title{
Novel coupled channel framework connecting quark model and lattice QCD: an investigation on near-threshold $D_s$ states}
\author{Zhi Yang}\email{zhiyang@uestc.edu.cn}
\uestc

\author{Guang-Juan Wang}\email{wgj@pku.edu.cn, corresponding author}
\asrc

\author{Jia-Jun Wu}\email{wujiajun@ucas.ac.cn, corresponding author}
\ucas

\author{Makoto Oka}\email{oka@post.j-parc.jp}
\asrc
\riken

\author{Shi-Lin Zhu}\email{zhusl@pku.edu.cn}
\pku

\date{\today}

\begin{abstract}
A novel framework is proposed to extract near-threshold resonant states from finite-volume energy levels of lattice QCD and is applied to elucidate structures of the positive parity $D_s$. The quark model, the quark-pair-creation mechanism and $D^{(*)}K$ interaction are incorporated into the Hamiltonian effective field theory. The bare $1^+$ $c\bar s$ states are almost purely given by the states with heavy-quark spin bases. The physical $D^*_{s0}(2317)$ and $D^*_{s1}(2460)$ are the mixtures of bare $c\bar s$ core and $D^{(*)}K$ component, while the $D^*_{s1}(2536)$ and $D^*_{s2}(2573)$ are almost dominated by bare $c\bar{s}$. Furthermore, our model reproduces the clear level crossing of the $D^*_{s1}(2536)$ with the scattering state at a finite volume.

\end{abstract}

\maketitle

Since first proposed by M. Gell-Mann~\cite{GellMann:1964nj} and G. Zweig~\cite{Zweig:1981pd}, the  quark model based on the valence quarks and anti-quarks has quite successfully explained the properties of the ground mesons and baryons~\cite{Eichten:1974af,Appelquist:1974yr,Richard:1983tc,Theberge:1982xs,Godfrey:1985xj, Capstick:1986bm,Isgur:1989vq}. 
However, the coupled-channel effects due to the hadronic loops are not taken into account in such conventional quark models, which is extremely important for near-threshold states~\cite{Thomas:1982kv, Thomas:1983fh, Ericson:1983um, Zhu:1998wy, Zhou:2011sp}. 
These missing effects lead to a gap between the prediction and observation in the experiment. 
For example, two lowest $S$-wave $c\bar{s}$ states, $D_s(1968,\,0^-)$ and $D_s^*(2112,\,1^-)$  are well described in the quark model, while the $P$-wave ones, $D_{s0}^*(2317)$ \cite{Aubert:2003fg} and $D^*_{s1}(2460)$~\cite{Besson:2003cp} which are close to the $D^{(*)}K$ thresholds, are both lighter than the quark model predictions.

Meanwhile, for $D_{s0}^*(2317)$ and $D^*_{s1}(2460)$, there exist various investigations, including quenched and unquenched $c\bar{s}$ quark models~\cite{Godfrey:1985xj,Dai:2003yg, Hwang:2004cd,Simonov:2004ar,Cheng:2014bca,Song:2015nia,Cheng:2017oqh,Luo:2021dvj,Zhou:2021uug,Alhakami:2016zqx}, molecule model~\cite{Kolomeitsev:2003ac,Szczepaniak:2003vy,Hofmann:2003je,vanBeveren:2003kd,
Barnes:2003dj,Gamermann:2006nm,Guo:2006rp, Guo:2006fu, Flynn:2007ki, Faessler:2007gv,Guo:2009ct,Xie:2010zza,Cleven:2010aw,Wu:2011yb,Guo:2015dha,Albaladejo:2016hae,Du:2017ttu,Guo:2018tjx,Albaladejo:2018mhb,Wu:2019vsy,Kong:2021ohg,Gregory:2021rgy,Wang:2012bu,Huang:2021fdt,Guo:2018tjx}, tetraquark model~\cite{Cheng:2003kg,Chen:2004dy,Dmitrasinovic:2005gc,Kim:2005gt,Zhang:2018mnm}, and $c\bar{s}$ plus tetraquark model~\cite{Terasaki:2003qa,Browder:2003fk,Maiani:2004vq,Dai:2006uz,Simonov:2004ar} (see reviews~\cite{Chen:2016spr,Dong:2017gaw,Guo:2017jvc,Yao:2020bxx}
for more details). 
However, their inner structures are still in a puzzle and the debating has never stopped until now. 
One biggest obstacle is the lack of experimental measurement for the scattering amplitude of the $D^{(*)}K \rightarrow  D^{(*)}K$ process. 
Fortunately, the lattice QCD simulation opens a new window to extract such information with the famous L\"uscher method~\cite{Luscher:1985dn,Luscher:1986pf, Luscher:1990ux}, which was introduced as a powerful technique linking the discrete energy levels from lattice QCD and the experimental observations, such as the scattering phase shifts and elasticities. 

Recently, several energy levels for the $D_s$ family were extracted  by lattice QCD simulation around  physical pion mass~\cite{Liu:2012zya,Mohler:2013rwa,Lang:2014yfa,Bali:2017pdv,Alexandrou:2019tmk}.
The energy levels below the thresholds can be recognized as the bound states, from which
the extracted masses of $D^*_{s0}(2317)$ and $D^*_{s1}(2460)$ are consistent with experimental measurements.
By using L\"uscher formalism~\cite{Luscher:1985dn,  Luscher:1986pf, Luscher:1990ux} and its developed equations (see review~\cite{Briceno:2017max}), the energy levels above the thresholds evolve into the scattering ones in the infinite volume.
The Hamiltonian effective field theory (HEFT)~\cite{Hall:2013qba, Wu:2014vma, Hall:2014uca, Liu:2015ktc} enables a quantitative examination of the lattice energy levels and scattering amplitudes in terms of hadronic degrees of freedom and their interactions. 
The two formalisms are equivalent if one ignores the exponential suppressed error ~\cite{Hall:2013qba,Wu:2014vma}.
Furthermore, the eigenvector from the Hamiltonian is helpful to probe the internal structure of coupled-channel system.
For instance, the property of $N^*(1535)$ was successfully determined~\cite{Liu:2015ktc}.

In this letter, we extend the HEFT by combining it with the quark model to study the nature of the mysterious near-threshold $D^*_{s0}(2317)$, $D^*_{s1}(2460)$, $D^*_{s1}(2536)$ and $D^*_{s2}(2573)$ states. 
The Hamiltonian contains the bare meson from the quark model, its coupling with the threshold channels described by quark-pair-creation (QPC) model~\cite{Micu:1968mk}, and the channel-channel interactions induced by exchanging light mesons.
These contributions, firstly together make a full phenomenological model to describe these $D_s$ states. 
This is an important development not only for understanding the physical picture of them but also a novel approach to study the nature of the near-threshold hadrons.  
The Godfrey-Isgur (GI) relativized quark model provided reasonably successful description for the spectra of low-lying mesons, from pion to bottomonium~\cite{Godfrey:1985xj}.
Nowadays, more experimental data are available for mesons, 
with which we improve the GI model parameters.
We use the masses of the well-established mesons that reside far away from the thresholds, 
in order to avoid possible mass shifts due to the coupled-channel effects.
With the updated parameters, the mass spectrum is better fitted to the experimental data than that in Ref.~\cite{Godfrey:1985xj}.

\begin{figure}[t]
\centering
\includegraphics[width=0.9\linewidth]{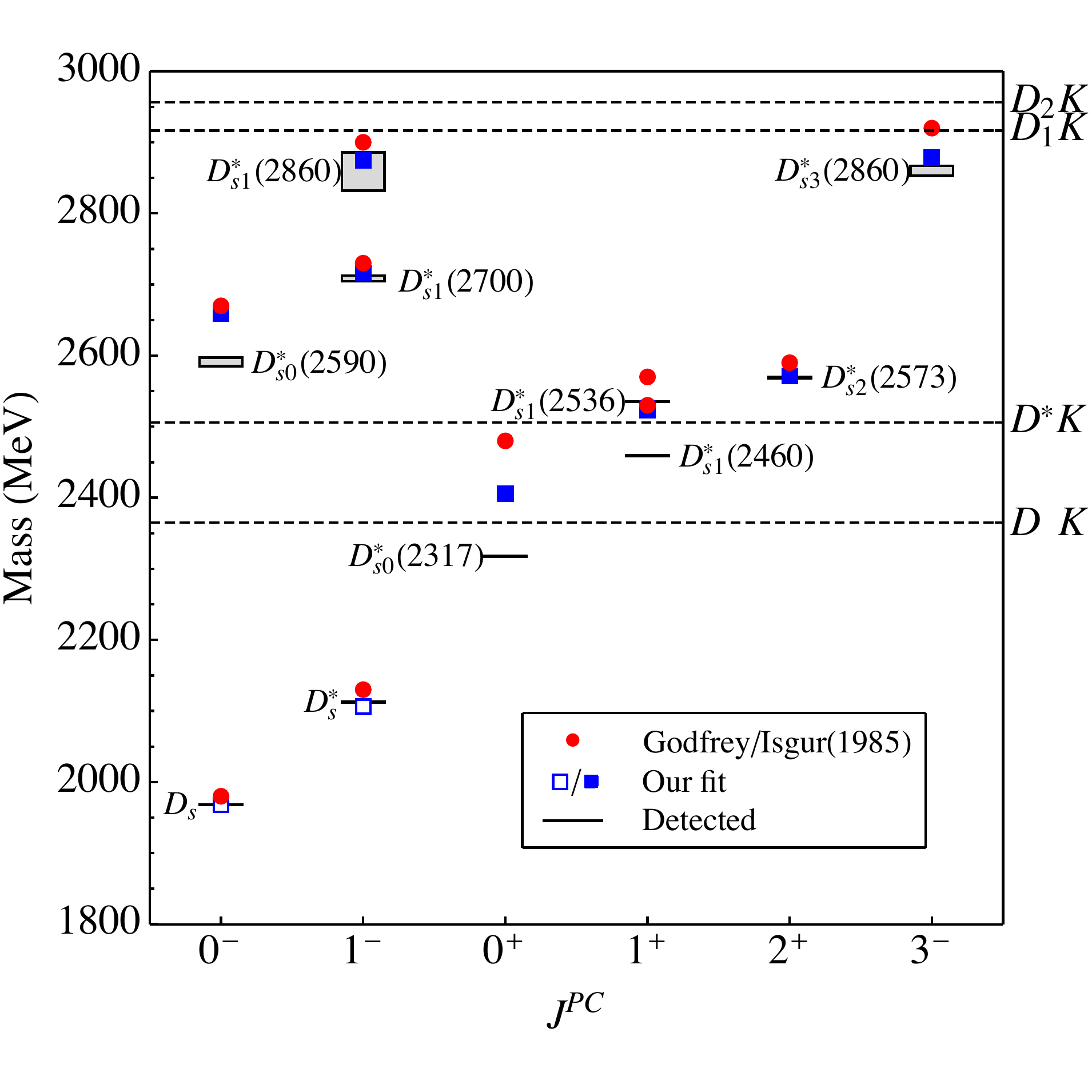}
\caption{
Mass spectrum of bare $c\bar s$ mesons within the relativized quark model. 
The circles and squares are the results predicted in Ref.~\mbox{\cite{Godfrey:1985xj} }\hspace{0pt} and our new fit, respectively.
In this sector, the two lowest lying $D^{(*)}_s$ states shown with open squares are used as input to constrain the quark model parameters.
The shaded areas correspond to the experimental masses and their uncertainties~\cite{Zyla:2020zbs, Aaij:2020voz}.}
\label{fig:Ds}
\end{figure}

We present the spectrum of $c\bar s$ mesons in the original GI model and the updated one in Fig.~\ref{fig:Ds}.
Even with the improved parameters, the masses of $D_{s0}^*(2317)$ and $D_{s1}^*(2460)$ are
significantly larger than the experimental data. 
Thus we believe that the coupled-channel effects are important to the two $D_s$’s due to the nearby 
$D^{(*)}\bar{K}$ thresholds.

The HEFT framework provides a multiple-component picture for a physical hadron. 
In the rest frame, the Hamiltonian reads,
\begin{equation} \label{eq:hamiltonian}
H=H_0+H_I\,,
\end{equation}
where the non-interacting Hamiltonian is,
\begin{equation}\label{eq:qmhamiltonian}
H_0=\sum_{B} |B\rangle\, m_{B}\,\langle B | + 
\sum_\alpha \int d^3 \vec{k} \,|\alpha(\vec{k})\rangle \,E_\alpha(\vec{k})\, \langle \alpha(\vec{k})|\,.
\end{equation}
Here $B$ denotes a bare $c\bar s$ core with the mass $m_{B}$ extracted from the GI model. 
The $\alpha$ represents the $D^{(*)}\bar{K}$ channels, and $E_\alpha(\vec{k})=\sqrt{m_K^2+\vec{k}^2}+\sqrt{m^2_{D^{(*)}}+\vec{k}^2}$ ($\vec{k}$ is the relative momentum) is the kinematic energy.
The $H_I=g+v$ is the energy independent interaction composed of two parts, the potential $g$ between the bare $c\bar s$ core and two-body channels $D^{(*)}K$, and the direct potential $v$ in the two-body channels.

The potential $g$ reads
 \begin{eqnarray}
&&g=\sum_{\alpha, B}\int d^3\vec{k} 
\left\{|\alpha(\vec{k})\rangle \,g_{\alpha\,B}(|\vec{k}|)\,\langle B |+h.c.
\right\}\,,
\end{eqnarray}
where $ g_{\alpha\,B}(|\vec{k}|)$
is obtained by the phenomenological QPC model \cite{LeYaouanc:1977fsz,Kokoski:1985is,Page:1995rh,Blundell:1996as,Ackleh:1996yt,Morel:2002vk, Ortega:2016mms} 
in which the bare $c\bar s$ core couples with the $D^{(*)}K$ channels through the creation of the light quark pair with the quantum number $J^{PC}=0^{++}$. 
Its explicit form is
 \begin{equation}
g_{\alpha\,B}(|\vec{k}|)=\gamma I_{\alpha\,B}(|\vec{k}|) e^{-\frac{\vec{k}^2}{2\Lambda^{\prime 2}}}\,,
\end{equation}
where $\gamma$ is a free parameter containing the creation probability of the quark-antiquark pair. 
The exponential form factor with the cutoff $\Lambda^{\prime}$ is introduced to truncate the hard vertices given by usual QPC model~\cite{Morel:2002vk, Ortega:2016mms}. 
The spatial transform factor $I_{\alpha\,B}(|\vec{k}|)$ is calculated with the exact wave functions obtained by our new fit. 

The potential $v$ in the two-body channels is defined as, 
\begin{eqnarray}
v=\sum_{\alpha,\,\beta}\int d^3\vec{k}\,d^3\vec{k}' 
\,|\alpha(\vec{k})\rangle\, V^L_{\alpha,\,\beta}(|\vec{k}|,\,|\vec{k}'|)\,\langle \beta(\vec{k}')|\,.
\end{eqnarray}
where  $V^L_{\alpha,\,\beta}(|\vec{k}|,\,|\vec{k}'|)$ is the $L$-wave potential between $\alpha$ and $\beta$ channels. 
Here we consider the $PP\rightarrow PP$ and $VP\rightarrow VP$ processes by exchanging light mesons, where $P$ and $V$ represent the $4\times 4$ pseudoscalar and vector meson matrices in the $SU(4)$ flavor symmetry, respectively.
Then, the $V^L_{\alpha,\,\beta}(|\vec{k}|,\,|\vec{k}'|)$  is straightforwardly obtained by the Lagrangian~\cite{Lin:1999ad,Oset:2009vf,Zhao:2014gqa} 
\begin{eqnarray}
\mathcal L &=& \mathcal L_{PPV}+\mathcal L_{VVV}\nonumber\\
&=&i g_v \text{Tr} (\partial^\mu P\,[P,V_\mu]\,)+ig_v\text{Tr} (\partial^\mu V^\nu\,[V_\mu,V_\nu]\,)\,,
\label{eq:lag}
\end{eqnarray}
where $g_v$ is the overall coupling constant.
To include the effects of the hadron structures, we introduce the form factors with a cutoff parameter $\Lambda$ for the interaction vertex, 
\begin{equation}
\left(\frac{\Lambda^{2}}{\Lambda^{2}+p_{f}^{2}}\right)^{2}\left(\frac{\Lambda^{2}}{\Lambda^{2}+p_{i}^{2}}\right)^{2}.\label{eq:ff}
\end{equation}

 For $D_s$ hadrons, we consider the bare $c\bar s$ cores from the GI model and the two possible coupled channels $D^{(*)}K$. 
 The coupling of the bare $c\bar s$ cores with the $D^{*}_s\pi$ or $D_s \gamma$ channels can be neglected, since these couplings arise from the isospin breaking interactions and electromagnetic ones. 
 Other possible strongly coupled channels are located far from the physical states and therefore not considered in this work, such as $D_s\eta$ for $D^*_{s0}(2317)$.
 We can construct three Hamiltonians for the physical $D_s$ states with the quantum numbers $J^P=$ $0^+$, $1^+$ and $2^+$, respectively. 
 The related bare $c\bar s$ cores and the $D^{(*)}K$ channels are shown in Table \ref{tab:mass}.

\begin{table}[t]
\caption{
The related bare $c\bar s$ cores ($B$)  and the $D^{(*)}K$ ($\alpha$) channels in the Hamiltonians of the physical $D_s$ states. 
The wave functions and mass spectrum (MeV) of the $B$ are shown. 
$\phi_s=|\frac{1}{2}_l\otimes \frac{1}{2}_h\rangle$ and  $\phi_d=|\frac{3}{2}_l\otimes \frac{1}{2}_h\rangle$ are  the heavy quark symmetry bases, where $h$ and $l$ are the heavy and light degrees of freedom, respectively.
The script $L$ in the last column denotes the orbital excitation in the $D^{(*)}K$ channels.
}
\label{tab:mass}
\begin{ruledtabular}
\begin{tabular}{rccccc}

&  &  $B(|^{2S+1}L_J\rangle)$                        & $B$(mass) & $\alpha$ & $L$ \\
\hline
 &  $D_{s0}^*(2317)$   & $|^3P_0\rangle$                              & 2405.9 & $DK$ & $S$  \\
 &  $D_{s1}^*(2460)$ & $0.68\,|^1P_1\rangle-0.74\,|^3P_1\rangle$    & 2511.5  & $D^*K$ & $S,\,D$ \\
       & & $=-0.99\,\phi_s
          +0.13\,\phi_d$  &  &      &\\
$   $    &  $D_{s1}^*(2536)$ & $-0.74\,|^1P_1\rangle-0.68\,|^3P_1\rangle$    & 2537.8 & $D^*K$  & $S,\,D$ \\
       && $=-0.13\,\phi_s
          -0.99\,\phi_d$  &  &      &\\
 &  $D_{s2}^*(2573)$   & $|^3P_2\rangle$                              &  2571.2 & $D^{(*)}K$ & $D$ \\

\end{tabular}
\end{ruledtabular} 
\end{table}

In the infinite volume, the scattering $T$-matrix between channels can be solved from the relativistic Lippmann-Schwinger equation~\cite{Matsuyama:2006rp,Wu:2012md,Wu:2014vma,Liu:2015ktc},
\begin{eqnarray}
T_{\alpha,\,\beta}(k,k';E)&=&{\mathcal V}_{\alpha,\,\beta}(k,k';E)+\sum_{\alpha'}\int q^2dq
\nonumber\\
&&\times \frac{ {\mathcal V}_{\alpha,\,\alpha'}(k,q;E)T_{\alpha,\,\beta}(q,k';E)}{E-E_{\alpha'}(q)+i\epsilon}\,,
\end{eqnarray}  
where the effective~potential ${\mathcal V}_{\alpha,\,\beta}(k,k';E)$ can be got from the interaction Hamiltonian,
\begin{equation}
{\mathcal V}_{\alpha,\,\beta}(k,k';E)=\sum_{B}\frac{g_{\alpha\,B}(k)\,g^*_{\beta\,B}(k')}{E-m_{B}}+V^L_{\alpha,\,\beta}(k,k')\,.
\end{equation}
The pole positions of bound states or resonances are obtained by searching for the poles of the $T$-matrix in the complex plane.

On the other hand, in a box with length $L$, the available momentum is integral multiples of the lowest non-trivial momentum $2\pi/L$ in any one dimension.
The Hamiltonian is translated into the discrete form featured by $n=n_x^2+n_y^2+n_z^2$ and the bare states.
The energy levels in the finite box correspond to the eigenvalues of the Hamiltonian matrix.  
Squares of the coefficients in the eigenvectors represent the probabilities $P(\alpha)$ ($\alpha =c\bar s, D^{(*)}K$) of the bare $c\bar{s}$ and $D^{(*)}K$ components~\cite{Wu:2014vma}.

In our model, there are four free parameters: the $\gamma$ and the cutoff $\Lambda^{\prime}$ in QPC model, the coupling constant $g_c$ ($g_c \propto g^2_v$) combing the $D^{(*)}D^{(*)}V$ and $KKV$ vertices and the cutoff $\Lambda$ in the $D^{(*)}K$ interactions. 
There are two groups of lattice data obtained using the pion mass $m_{\pi}=150, 156$ MeV for the  $J^P=0^+, 1^+$ $D_s$ sectors in Refs.~\cite{Lang:2014yfa,Bali:2017pdv}. 
The chiral extrapolation therefore is not considered in this work. 
We perform a simultaneous fit of two lattice data sets in the $J^P=0^+$ (left) and $1^+$ (middle) sectors as shown in Fig.~\ref{fig:fitspec}. 
The cutoff $\Lambda$ is taken as 1 GeV, noting that its dependence can be absorbed by the renormalization of the interaction kernel (details are in the Supplemental Material), and then the other parameters are fitted as
\be
g_c = 4.2^{+2.2}_{-3.1}\;,
\Lambda^{\prime} = 0.323^{+0.033}_{-0.031}\;\text{GeV},
\gamma = 10.3^{+1.1}_{-1.0}
\ee
with $\chi^2/\text{dof}=0.95$. 
The parameters are roughly consistent with other phenomenological investigations~\cite{Godfrey:2015dia,Shen:2019evi}.
With above parameters determined by the lattice QCD data, we obtain the pole masses of the $T$-matrix as listed in Table~\ref{tab:fitmass}, which agree with the experimental data.

For the $J^P=0^+$ case, one bare $c\bar s$ core and the $S$-wave $DK$ channel are included for the $D^*_{s0}(2317)$ as shown in Table \ref{tab:mass}.
In Table~\ref{tab:fitmass}, the pole position is located at $2338.9$ MeV in the first Riemann-sheet of $DK$ channel.  
Because of the larger input data from lattice QCD which is likely to be due to discretization effects as was pointed out in Refs.~\cite{Bali:2017pdv, Lang:2014yfa}, the computed mass is around $21$ MeV larger than experimental data. However, this  small discrepancy is not expected to change our main conclusions. 
By analyzing  the eigenvector, the bare $c\bar s$ core in $D^*_{s0}(2317)$ occupies around $32.0\%$ at $L=4.57$ fm, while $DK$ component accounts for around $68.0\%$. 
This is consistent with the result $P(DK)=(67\pm14)\%$ from Ref.~\cite{Torres:2014vna}.
Despite the probability not being an observable, it will be related to the decay patterns of the $P$-wave $D^*_s$'s due to the different strong and radiative decays of the $c\bar{s}$ and $DK$ components~\cite{Godfrey:2003kg,Mehen:2004uj,Wang:2006fg,Faessler:2007gv,Guo:2008gp,Cleven:2014oka,
Fajfer:2015zma,Fu:2021wde}.
Here, the $P(\alpha)$ shows that the two components are significant and essential for the $D_{s0}^*(2317)$ state.
Furthermore, we have performed the fit without coupling to  the bare $c\bar s$, and found that the Hamiltonian matrix cannot describe the Lattice data of $D_{s0}^* (2317)$ only with the $D^{(*)}K$ component. This proves that the bare core is indispensable in the formation of the physical state.

For the $J^P=1^+$ case, it includes two bare $c\bar s$ cores for $D^*_{s1}(2460)$ and $D^*_{s1}(2536)$, and two $D^*K$ channels with $S$- and $D$-wave orbital excitations as shown in Table~\ref{tab:mass}. 
These two bare $c\bar s$ cores in the quark model lie close to the $D^*K$ channels as illustrated in Fig.~\ref{fig:Ds}. 
However, they are dominated by the $|\frac{1}{2}_l\otimes \frac{1}{2}_h\rangle$ and $|\frac{1}{2}_l\otimes \frac{3}{2}_h\rangle$ components, respectively. 
Within a good heavy quark symmetry, the lighter and heavier bare $c\bar s$ cores mainly couple with the $S$- and $D$-wave $D^*K$ channels, respectively. 
In the middle panel of Fig.~\ref{fig:fitspec}, the lighter bare $c\bar s$ core has a significant mass shift due to the $S$-wave interaction and becomes the lowest eigenstate corresponding to the $D^*_{s1}(2460)$ state which is the mixture of the bare $c\bar s$ core and $D^{*}K$ component with $P(D^*K)\approx47.6\%$ as shown in Table~\ref{tab:fitmass}. 
In contrast, the $D$-wave interaction around the threshold is significantly suppressed at $\mathcal O(k^2)$ compared with the $S$-wave one.
Therefore, the energy level of $D_{s1}^*(2536)$ almost keeps stable, and its bare $c\bar s$ core dominates with $P(c\bar s)\approx98.2\%$.

Meanwhile, a special crossing happens above the $D^*K$ threshold in $1^+$ sector of Fig.~\ref{fig:fitspec} around $L=3.5$ fm.
The dropping line is dominated by the lowest excited  $D^*K$ channel with the  kinematic energy depending on $(\frac{2\pi}{L})^2$,
while the flat line represents the $D^*_{s1}(2536)$ state.
With $L=3.5$ fm, the energy levels of the lowest excited  $D^*K$ channel and the $D^*_{s1}(2536)$ state are nearly degenerate, which leads to the crossing.
This crossing is well proved by the lattice data. 
Above the $D^*K$ threshold, the two data points are almost pinched at $L=3.4$ fm while others are distinguishable at $L=2.9,\, 4.8$ fm. 
One notes that the lattice data close to the flat line were extracted mainly by the $c\bar s$ operator~\cite{Lang:2014yfa,Bali:2017pdv}, which is completely consistent with our picture where the majority of the $D^*_{s1}(2536)$ is the bare $c\bar s$ core. 
%

\begin{figure*}[!htp]
\centering
\includegraphics[width=0.8\linewidth]{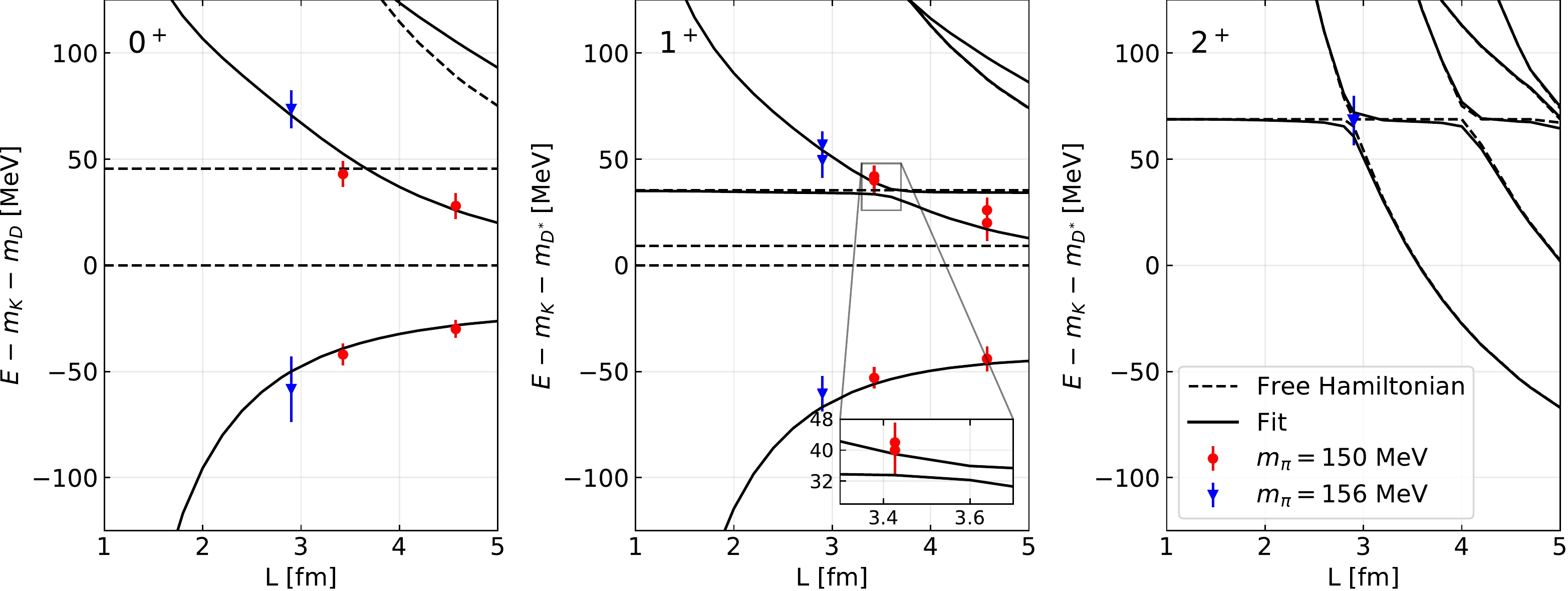}
\caption{
The fitted binding energy dependence of the length $L$ for the $D^*_{s0}(2317)$ (left), the $D^*_{s1}(2460/2536)$ (middle) states with the pion mass $m_{\pi}=150$ MeV~\cite{Bali:2017pdv} and $m_{\pi}=156$ MeV~\cite{Lang:2014yfa}. 
The comparison of the lattice binding energies and our predicted ones for $D^*_{s2}(2573)$ is shown in the right panel.
The black curves are the results using  finite-volume Hamiltonian, while the dashed lines represent the masses of the bare $c\bar s$ cores and $D^{(*)}K$ thresholds obtained with the free Hamiltonian $H_0$.}
\label{fig:fitspec}
\end{figure*}

\begin{table}[t]
\caption{
The comparison of $D_s$ pole masses (MeV) (ours) with the experimental results (exp). 
The script $P(c\bar{s})$ represents the content of the bare $c\bar s$ cores in the $D_s$ states at $L=4.57$ fm. }
\label{tab:fitmass}
\begin{ruledtabular}
\begin{tabular}{cccc}
        & $P(c\bar{s})[\%]$  & ours  & exp   \\
\hline
$D^*_{s0}(2317)$   &  $32.0^{+5.2}_{-3.9}$  & $2338.9^{+2.1}_{-2.7}$ &  $2317.8\pm0.5$    \\
$D^*_{s1}(2460)$   &  $52.4^{+5.1}_{-3.8}$  & $2459.4^{+2.9}_{-3.0}$ &  $2459.5\pm0.6   $ \\
$D^*_{s1}(2536)$   &  $98.2^{+0.1}_{-0.2}$  & $2536.6^{+0.3}_{-0.5}$ &  $2535.11\pm0.06$  \\
\hline
$D^*_{s2}(2573)$   &  $95.9^{+1.0}_{-1.5}$  & $2570.2^{+0.4}_{-0.8}$ &  $2569.1\pm0.8$  \\
\end{tabular}
\end{ruledtabular} 
\end{table}

To verify our model, we give the prediction for the $D^*_{s2}(2573)$ with the fitted parameters. 
Here, the Hamiltonian matrix includes one bare $c\bar s$ core and two $D$-wave channels, $DK$ and $D^*K$. 
For $J^P=2^+$ case, the energy levels are shown in right panel of Fig.~\ref{fig:fitspec}.
Because of the weak $D$-wave interaction, the $D^*_{s2}(2573)$ is almost a pure $c\bar s$ state with $P(c\bar{s})\approx 95.9\%$.


In summary, we have incorporated the quark model, the QPC model, and the coupled channel unitary approach into the HEFT. Then, it is connected to the lattice QCD to investigate the lowest four $D_s$ states with $J^P=0^+/1^+/2^+$ for the first time. 
By fitting the recent energy levels on lattice QCD for the three lowest $0^+$ and $1^+$ $D_s$ states, we successfully build a systematical model for the $D^*_{s0}(2317)$, $D^*_{s1}(2460)$,  $D^*_{s1}(2536)$, and $D^*_{s2}(2573)$ states. %
The obtained pole masses are well consistent with experimental data. 
Moreover, the model provides a clear physical picture for the $D_{s}$ family with positive parity.
The $D^*_{s0}(2317)$ and $D^*_{s1}(2460)$ states have the mass shifts by tens of MeV because of the coupled-channel effects with the $S$-wave $DK$ and $D^{*}K$ channels, respectively.
They are the mixtures of the bare $c\bar s$ core and $D^{(*)}K$ component, while the $D^*_{s1}(2536)$ and $D^*_{s2}(2573)$ states are almost pure $c\bar{s}$ mesons because of the kinematically suppressed $D$-wave coupling.

In addition, it is worth emphasizing that the bare state plays an extremely important role to form the physical $D^*_{s0}(2317)$ in our model. 
Further investigation can be done in lattice QCD to examine this conclusion. 
With increasing pion mass ($m_\pi$), the mass of the bare $c\bar s$ state will be almost stable, similar to that of $D_s(1968)$ in the lattice simulation. 
However, the $DK$ component contains light valence quarks. Its mass will keep increasing with larger $m_\pi$~\cite{Bali:2017pdv}. 
If $D^*_{s0}(2317)$ is mainly a $c\bar s$ core, the corresponding energy level will finally approach the mass of the bare $c\bar s$ state although it may increase at first. 
Otherwise, it will keep increasing~\cite{Du:2017ttu}.
There exists very limited data from different lattice groups so far~\cite{Bali:2017pdv,Cheung:2020mql}. 
We strongly suggest lattice QCD groups to make a systematical investigation regarding the mass dependence of the $D^*_{s0}(2317)$ on $m_\pi$.

Furthermore, the model about $D_s$ family should be helpful in the relevant analysis of experimental processes, such as $B_{s}/B \to D^{(*)}D^{(*)}K$ or $D^{(*)}KK$.  
In these decays, the $D^{(*)}K \to D^{(*)}K$ amplitude cannot be fully obtained because of the unknown vertex related to the $B_s/B$ state.  
A theoretically motivated model for the parameterization of $D^{(*)}K \to D^{(*)}K$ amplitude is necessary.

Finally, the HEFT has built a bridge among the phenomenological models, the patterns of the lattice QCD data, and experimental data. 
This formalism can be extended to study other states lying close to the two-meson thresholds, for instance, the XYZ exotic states.
Such investigation can help disentangle their nature and deepen our understanding of the nonperturbative QCD in the future.

\begin{acknowledgments}

We thank useful discussions and valuable comments from Xiang Liu, Zhihui Guo, Lisheng Geng, Wei Wang, Liuming Liu, Fengkun Guo, Bing-Song Zou, Ross D. Young and James M. Zanotti.
We thank for Philipp Gubler's careful reading and useful suggestions. 
This work is partly supported by the National Natural Science Foundation of China (NSFC) under Grants Nos. 11847301 (Z.Y.), 
and by the Fundamental Research Funds for the Central Universities under Grant No. 2019CDJDWL0005 (Z.Y.), 
and by the supported by JSPS KAKENHI under Grant No. 20F20026(G.J.W.), 
and by the Fundamental Research Funds for the Central Universities (J.J.W.), 
and by the National Key R$\&$D Program of China under Contract No. 2020YFA0406400 (J.J.W.),
and by the JSPS KAKENHI under Grant Nos.~19H05159, 20K03959, and 21H00132 (M.O.),
and by the National Natural Science Foundation of China under Grant Nos.11975033 and 12070131001 (S.L.Z.).

\end{acknowledgments}
\bibliography{Ds.bib}

\onecolumngrid
\newpage
\section*{Supplemental Material}

\subsection{The investigation of $\Lambda$ dependence}

The typical value of $\Lambda$ is  usually in the range $\Lambda \sim 0.9\pm 0.2~\text{GeV}$.
Here, we provide  two additional new fits with $\Lambda = 0.8 $ and
$1.2$ GeV.
The final results are shown in Table~\ref{tab:fitmass} and Fig.~\ref{fig:fitspec}. The three sets of
results are similar to each other, which implies that the $\Lambda$
dependence can be absorbed by the renormalization of the interaction
kernel.
And the final results are almost independent on the choice of
$\Lambda$ in the widely used region.
\begin{figure*}[!htp]
\centering
\includegraphics[width=0.8\linewidth]{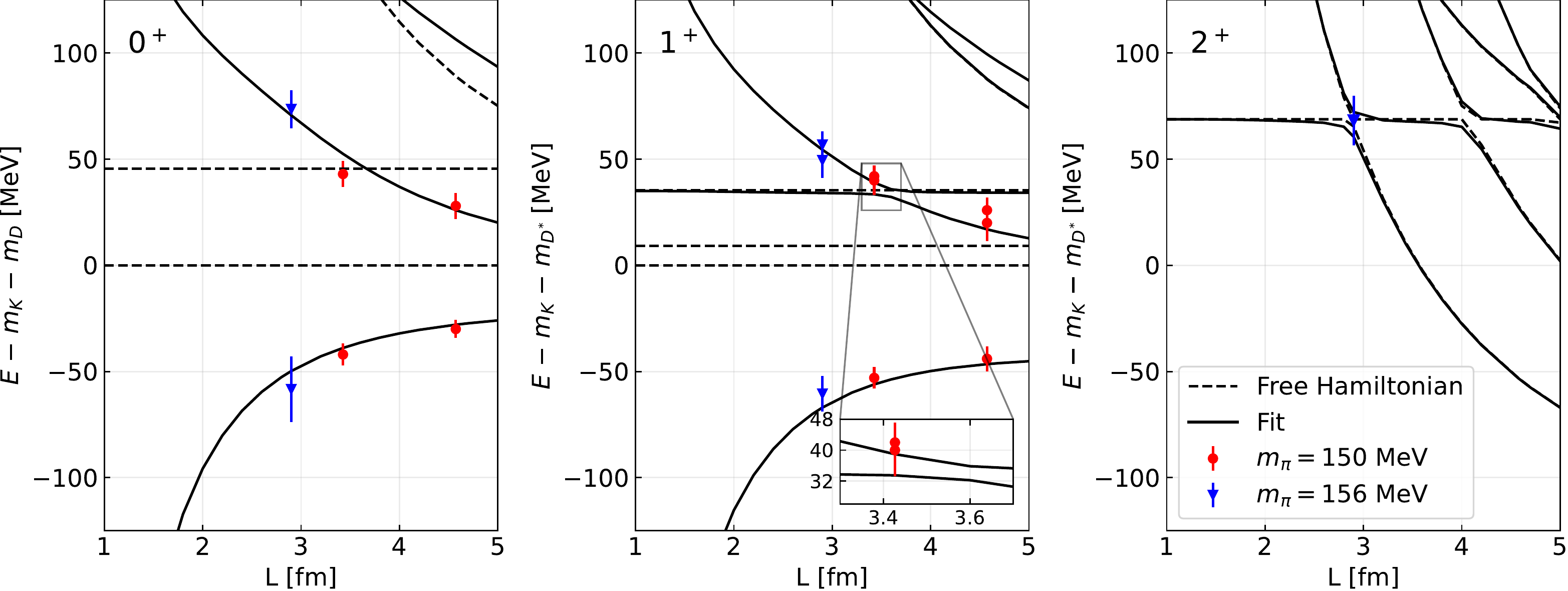}\vspace{0.5cm}
\includegraphics[width=0.8\linewidth]{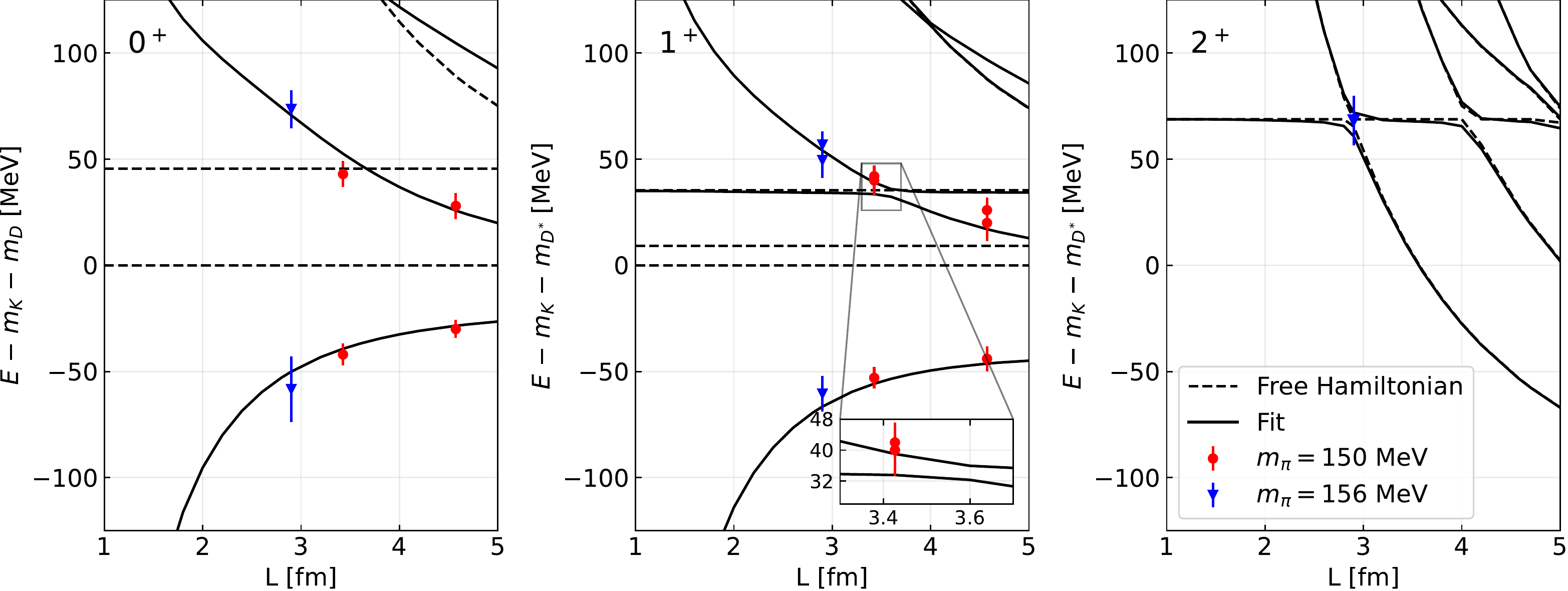}
\caption{The fitted binding energy dependence of the
length $L$ for the $D^*_{s0}(2317)$ (left), the
$D^*_{s1}(2460/2536)$ (middle) states with the pion mass
$m_{\pi}=150$ MeV~[69] and $m_{\pi}=156$ MeV~[68] for $\Lambda =
0.8,1.2$ GeV from up to down in order.} \label{fig:fitspec}
\end{figure*}
\\
\begin{table}[t]
\caption{The content of the bare $c\bar s$ cores in
the $D_s$ states $P(c\bar{s})$ for three different $\Lambda$.
}\label{tab:fitmass}
\begin{ruledtabular}
\begin{tabular}{cccc}
        & $\Lambda=1.0 \,[\%]$  &  $\Lambda=0.8\,[\%]$  &  $\Lambda=1.2\,[\%]$   \vspace{0.1cm}\\
\hline
$D^*_{s0}(2317)$   &  $32.0^{+5.2}_{-3.9}$  & $32.5^{+4.7}_{-3.6}$ &  $31.6^{+5.9}_{-3.5}$ \vspace{0.1cm}   \\
$D^*_{s1}(2460)$   &  $52.4^{+5.1}_{-3.8}$  & $53.0^{+4.5}_{-3.9}$ &  $51.9^{+5.9}_{-3.9}$ \vspace{0.1cm} \\
$D^*_{s1}(2536)$   &  $98.2^{+0.1}_{-0.2}$  & $98.2^{+0.1}_{-0.2}$ &  $98.2^{+0.1}_{-0.2}$ \vspace{0.1cm} \\
\hline
$D^*_{s2}(2573)$   &  $95.9^{+1.0}_{-1.5}$  & $95.7^{+0.9}_{-1.3}$ &  $96.0^{+0.8}_{-1.7}$ \vspace{0.1cm}  \\
\end{tabular}
\end{ruledtabular}
\end{table}

\end{document}